\documentclass[twocolumn,prb]{revtex4}
\usepackage{graphicx}
\usepackage{dcolumn}
\usepackage{bm}

\usepackage{graphics}

\begin{document}

\title{Coulomb gap and variable range
hopping in a pinned Wigner crystal}

\author{B.~I.~Shklovskii}

\affiliation{William I. Fine Theoretical Physics Institute,
University of Minnesota, Minneapolis, Minnesota 55455}

\date{\today}
\begin{abstract}

It is shown that pinning of the electron Wigner crystal by a small
concentration of charged impurities creates the finite density of
charged localized states near the Fermi level. In the case of
residual impurities in the spacer this density of states is
related to nonlinear screening of a close acceptor by a Wigner
crystal vacancy. On the other hand, intentional doping by a remote
layer of donors is a source of a long range potential, which
generates dislocations in Wigner crystal. Dislocations in turn
create charged localized states near the Fermi level. In both
cases Coulomb interaction of localized states leads to the soft
Coulomb gap and ES variable range hopping at low enough
temperatures.

\end{abstract}
\maketitle

The growth of GaAs-heterostructures mobilities has made possible
to study collective properties of the hole gas with a very small
concentration. The average distance between holes $r_s$,
calculated in units of the hole Bohr radius,  has reached 80 and
exceeded the theoretically predicted freezing point of the hole
liquid, $r_s =38$, where it should become the Wigner crystal. The
transport of such low density systems with and without magnetic
field was studied recently. If a strong magnetic field applied
along the hole plane the low temperature conductivity of the
Wigner-crystal-like phase seems to be related to the variable
range hopping (VRH)~\cite{Noh}. Thus, it seems timely to discuss
ohmic transport of the Wigner crystal. Of course, any Wigner
crystal is always pinned by impurities and can not slide as a
whole in a small electric field. Interstitials and vacancies of
the Wigner crystal provide conductivity but they have large
activation energies and freeze out at low temperatures. On the
other hand, the VRH conductivity requires nonzero density of
states of charge excitations near the Fermi level, both in the
case of Mott law or Efros-Shklovskii (ES) law. If pinning centers
produce such "bare" density of states then the long range Coulomb
interaction creates the Coulomb gap around the Fermi level and
leads to the ES variable range hopping. In this paper, I study the
origin of the bare density of states of the pinned Wigner crystal,
the Coulomb gap on its background and the resulting VRH
conductivity at low temperatures.

Let me start from  a clean electron Wigner crystal on uniform
positive background. Consider an additional electron added to the
perfect Wigner crystal and let the lattice relax locally. This
addition costs the interstitial energy $0.14 e^{2}/\kappa a$,
where $a$ is the lattice constant or the optimal triangular
lattice, $\kappa$ is the dielectric constant of
GaAs~\cite{Fisher}. On the other hand, the energy cost of the
extraction of an electron and the corresponding relaxation around
the vacancy is equal to the vacancy energy $0.26 e^{2}/\kappa a$.
Thus, the density of states of relaxed excitations (electronic
polarons~\cite{ES84}), consists of two delta-like peaks at
energies $-0.26 e^{2}/\kappa a$ and $0.14 e^{2}/\kappa a$. The
lower peak corresponds to occupied states, the upper one contains
empty states. Between them  there is the hard (completely empty)
gap of the width $0.40 e^{2}/\kappa a$. The Fermi level of
electronic polarons (the zero of energy) is situated in this gap,
therefore the density of states at the Fermi level is zero.

In typical samples the Wigner crystal is pinned by charged
impurities: donor and acceptors. Even if there is no intentional
doping and the two-dimensional electrons gas (2DEG) is created in
a heterostructure by a distant metallic gate, there is always a
small concentration of residual donors and acceptors. Many high
mobility heterostructures are intentionally doped by a donor
layer, which is separated from 2DEG by the spacer with a large
width $s \gg a$. This is an additional source of disorder. In the
beginning of this paper we neglect this source and deal with
residual impurities only. We return to the role of doped layer in
the second part of this paper.

Let me argue that close residual donors and acceptors create
localized states at the Fermi level. I use theory of pinning of
the Wigner crystal by charge impurities developed in
Ref.~\onlinecite{Rouzin}. As in that paper I talk about Wigner
crystal of electrons and concentrate on close residual acceptors
which according to Ref.~\onlinecite{Rouzin} produce stronger
pinning than donors. (For the hole crystal the role of acceptors
is of course played by donors.) Because the spacer width, $s$, and
the lattice constant of the Wigner crystal, $a$, are much larger
than the lattice constant of AlGaAs the authors of
Ref.~\onlinecite{Rouzin} assumed that acceptors are randomly
distributed in space with three-dimensional concentration $N$.
Each acceptor is negative because it has captured one electron
from 2DEG.

It was found~\cite{Rouzin} that effect of an acceptor dramatically
depends on its distance, $d$, from the plane of the Wigner
crystal. If $d$ is larger than $0.68a$ the electron crystal
adjusts to acceptor in such a way that in the ground state an
interstitial position of the crystal is exactly above the
acceptor. In this case, interaction of the acceptor with the
Wigner crystal can be calculated in the elastic approximation. On
the other hand, when $d$ is smaller than $0.68a$ acceptor creates
a vacancy in the crystal, which positions itself right above the
acceptor. This means that the close acceptor effectively builds in
the crystal cite pushing away to infinity its electron. We can
consider the latter acceptor as an empty localized state for
electron, while the former acceptor can be considered as a
localized state occupied by an electron. Their energies are equal
at $d =0.68a$, so that all acceptors with $d > 0.68a$ are occupied
and all acceptor with $d < 0.68a$ are empty. Thus, acceptors with
$d = 0.68a$ are at the Fermi level and the bare density of states
near the Fermi level
\begin{equation}
g_B \sim \frac{Na}{e^{2}/\kappa a}, \label{gB}
\end{equation}
where $Na$ is an estimate of the two-dimensional concentration of
acceptors located within distance $a$ (only they provide
substantial pinning) and $e^{2}/\kappa a$ is the characteristic
energy of their pinning. The nonzero density of states $g_B$ makes
possible VRH conductivity of the pinned Wigner crystal. For
example, an electron from an acceptor with $d = 0.68a + 0$ can hop
to a distant one with $d = 0.68a - 0$.

The long range Coulomb interaction between localized states leads
to the soft Coulomb gap at the Fermi level~\cite{ES}. Shape of the
Coulomb gap depends on the interaction, $U(r)$, of the hopping
electron in the final state of the hop and the empty place it has
left behind (excitonic term)~\cite{ES,ES84}. If the interaction
can be described by the standard Coulomb law $U_{C}(r)= -
e^{2}/\kappa r$, where $r$ is the two-dimensional distance between
initial and final state, then the Coulomb gap has the standard for
two-dimensional case form
\begin{equation}
g(E) = \frac{2}{\pi} \frac{|E|\kappa^{2}}{e^4}. \label{CG}
\end{equation}
The attraction potential $U(r)$ generally speaking is different
from $U_{C}(r)$ due to screening of the Coulomb interaction by
elastic deformations of the Wigner crystal or in other words by
its polarization. In order to calculate Fourier image $U(q)$ of
$U(r)$ we have to introduce the Larkin length $L$ at which
acceptors destroy the long range order of the crystal. In the
two-dimensional Wigner crystal pinning by close acceptors is so
strong that this length is close to the average distance between
them, $L \sim (Na)^{-1/2}$. Using $L$, we can write
\begin{equation}
U(q)= 2\pi/q\kappa \epsilon(q), \label{Uq}
\end{equation}
where
\begin{equation}
\epsilon(q)= 1 + \frac{2}{qr_D}\frac{q^{2}}{q^{2} - L^{-2}}.
\label{epsilonq}
\end{equation}
is given by interpolation between large and small $q$ cases.  Here
$r_D\sim -0.32a $ is the linear Debye screening radius of the
Wigner crystal. (The asymptotics of Eq.~(\ref{epsilonq}) at large
and small $q$ are similar to those of the dielectric constant of a
two-dimensional electron gas in a strong magnetic
field~\cite{Kukushkin,Aleiner,Koulakov}, where the Larmour radius
plays the role of $L$.)

In the real space the Fourier image of Eq.~(\ref{epsilonq})
results in the Coulomb potential $U(r) = U_{C}(r)$ only at $r \gg
L^{2}/a = 1/Na^2$. One can say that by these distances all
electric lines of a probe charge located in the reference point in
the plane of Wigner crystal leave the plane. This leads to the
standard Coulomb gap Eq.~(\ref{CG}), but only at energies $|E| <
e^2 Na^{2}/\kappa$. At smaller distances $r \ll 1/Na^2$ potential
$U(r)$ is weaker than $U_{C}(r)$ and grows only logarithmically
with decreasing $r$. But this does not lead to faster
(exponential) growth of the density of states at $|E| \sim e^2
Na^{2}/\kappa$ as it would if $g_B$ were very large. The reason is
that at energy $|E| \sim e^2 Na^{2}/\kappa$ the Coulomb gap
density of states, Eq.~(\ref{CG}), already reaches the bare
density of states, $g_B$, given by Eq.~(\ref{gB}). The Coulomb gap
is just a depletion of the density of states on the background of
the bare density. Therefore, the density of states in the Coulomb
gap can not be larger than $g_B$~\cite{ES84}. This means that at
$|E| \gg e^2 Na^{2}/\kappa$ the density of states saturates at the
level of Eq.~(\ref{gB}).

The role of distance $L^{2}/a = 1/Na^2$ can be interpreted in
another way. When we add an electron to the clean Wigner crystal
it makes an interstitial but its charge spreads to infinity. In
the a pinned Wigner crystal an added charge is smeared in the disc
of the finite radius $R$. We can find $R$ minimizing the sum of
the Coulomb energy $e^{2}/\kappa r$ of the disc of radius $r$ and
the shear energy $\mu (u/L)^{2} r^{2}$ necessary for the disc
dilatation by the area $a^2$. Here $\mu \sim e^{2}/a^{3}$ is the
shear modulus and $u \sim a^{2}/r$ is the necessary displacement.
It is important that the characteristic distance of the variation
of the shear displacement (which percolates between strongly
pinning acceptors) is the average distance between close acceptors
$L \sim (Na)^{-1/2}$. Minimizing this sum we find that $R =
L^{2}/a = 1/Na^2$. This again tells us that at distances larger
than $L^{2}/a = 1/Na^2$ we deal with the unscreened Coulomb
potential.

At low enough temperatures the Coulomb gap leads to
Efros-Shklovskii(ES) law of the temperature dependence of the variable
range hopping~\cite{ES,ES84}
\begin{equation}
\sigma = \sigma_0 \exp[-(T_{ES}/T)^{1/2}]. \label{DC}
\end{equation}
Here $\sigma_0$ is a prefactor, which has only an algebraic
$T$-dependence and
\begin{equation}
T_{ES} = C e^2 / \kappa \xi, \label{TES}
\end{equation}
where $C$ is a numerical coefficient close to 6 and $\xi$ is the
localization length for electron tunnelling with energy close to
the Fermi level in the Wigner crystal. We will discuss the value
of $\xi$ in the end of the paper.

When with increasing temperature the width of the band of energies
used for ES hopping $(T T_{ES})^{1/2}$ reaches
$e^{2}Na^{2}/\kappa$ ES law is replaced by the Mott conductivity
\begin{equation}
\sigma = \sigma_0 \exp[-(T_{M}/T)^{1/3}], \label{DCM}
\end{equation}
where $T_M = C_{M} / (g_{B}\xi^2)$. Transition from the ES law to
the Mott law happens at $T \sim T_{ES}^{3}/T_{M}^2 \sim
T_{ES}(Na^2\xi)^{2}$. Thus, while ES law does not depend on the
acceptor concentration $N$ the range of ES law shrinks when
concentration of acceptors decreases.

This example emphasizes universality of the Coulomb gap and ES
law. The Coulomb gap was derived for lightly doped semiconductors,
where disorder is as strong as interactions~\cite{ES}. Later
Efros~\cite{Efros} suggested a model, where electrons are located
on sites  of a square lattice with +1/2 charges, the number of
sites being twice larger than number of electrons. Random energies
of sites are uniformly distributed in the band $A$ (they are
measured in units of Coulomb interaction of two electrons on
nearest sites). In this model  the Coulomb gap does not survive in
the small disorder case $A << 1$ because in this case in spite of
small disorder positive charges of empty sites and negative
charges of occupied sites alternate in the perfect NaCl-like
order. One could, therefore, say that the Coulomb gap is the
property of strong disorder only. The above example of the Wigner
crystal on the uniform background pinned by rare strong impurities
however shows that such impression may be misleading. Even in the
case when concentration of acceptors $N$ is small (one can call it
the weak disorder case) the Coulomb gap and ES law survive.

One can interpret what happens with the Wigner crystal in the
terms of the Efros model. At strong disorder when $Na^3 \sim 1$
our model is close to the Efros model with $A \sim 1$. When
parameter $Na^3$ becomes much smaller than unity and the disorder
becomes weak, the energy scatter of states created by impurities
stays at the level of $e^2/\kappa a$, while the long Coulomb
interaction between these states becomes much smaller, of the
order of $e^{2}(Na)^{1/2}\\kappa$. Introducing a two-dimensional
random lattice with cites occupied by acceptors we can come to a
renormalized Efros model with $A \sim 1/(Na^3)^{1/2} \gg 1$. This
brings us again to the Coulomb gap at small energies.

Until now we talked about the role of residual acceptors. Let us
qualitatively discuss intentionally doped heterojunctions where
donors are situated in the narrow layer parallel to the plane of
the 2DEG (delta-doping)  at a large distance $s \gg a$ from 2DEG.
Random distribution of donors in this layer creates fluctuations
of their potential of all wavelengths, but only harmonics of the
random potential with wavelengths larger than $s$ reach  2DEG. Let
us assume that the two-dimensional concentration of charged donors
in the doped layer is $N_{2}$. In principle $N_{2}$ can be
different from the two-dimensional concentration of electrons in
the Wigner crystal $n$ (2DEG can be compensated by acceptors or
created by illumination). If $N_{2} \ll n$ the Wigner crystal
screens external potential by small, purely elastic deformations.
Due to these deformations energies of an interstitial and a
vacancy depend on a coordinates. Therefore, both delta-like peaks
of the density of states are somewhat smeared. Their width,
however, is much smaller than the hard gap between them.  Thus,
the hard gap is preserved and no states at the Fermi level appear.
When the concentration of donors, $N_{2}$, grows, the amplitude
$u$ of displacements of electrons in the Wigner crystal with
wavelength of the order $s$ reaches the lattice constant of the
crystal $a$. Such large deformations resolve themselves by
creation of dislocations~\cite{FisherF,Fertig}, because the energy
price of the dislocation core becomes smaller than an elastic
energy, which is eliminated by the dislocation. This happens when
the donor concentration, $N_2$, is of the order of electron
concentration, $n$. Indeed, if we cover the layer of donors by
squares with the side $s$ the typical fluctuation of number of
charges in a square is equal $(N_{2}s^2)^{1/2}$. Potential of the
charge of this fluctuation reaches 2DEG practically without
compensation by oppositely charged fluctuations in neighboring
squares. As a result at $n =N_{2}$ 2DEG has to provide
$(Ns^2)^{1/2} = s/a$ new electrons to screen random potential in a
square. This may be done by an additional raw of electrons in the
square or, in other words, by two dislocations. Thus approximately
one dislocation appears in a square with a side $s$. This picture
is actually a simple visualization of the Larkin domain. Below we
concentrate on the case $N \sim n$. In this case, the length $s$
plays the role of the Larkin length $L$.

An isolated dislocation brings the electronic polaron state right
to the Fermi level. Indeed, if we add an electron from the Fermi
level to the end of an additional raw terminated by a dislocation
and let the rest of electrons relax, the dislocation just moves
along this additional raw by one lattice constant $a$ and the
energy remains unchanged. Similarly, if we extract an electron
from the end of the raw to the Fermi level, the dislocation moves
in the opposite direction and the energy does not change. This
means that if we neglect the interaction of dislocation with long
range ptential of donors and interaction between
dislocations they
create a delta peak of the density of states right at the Fermi
level. This peak is normalized on the concentration of
dislocations. The Fermi level is pinned in the middle of this
peak.

In the long range fluctuating potential of donors the peak of the
density of dislocation states is smeared, because dislocations
strongly interact with the gradient of potential. The peak is
smeared also due to the interaction between the dislocations. When
an additional electron is absorbed by a dislocation and, as a
result, the dislocation moves by one lattice site, the logarithmic
interaction with other dislocation changes.

To understand the role of Coulomb interaction
in the density of states of the pinned Wigner crystal one has to
concentrate on the fate of the charge
of an added electron. In the clean
Wigner crystal, if we have a single
dislocation and move it by one lattice constant
adding a new electron, the charge of this electron spreads to
infinity, leaving the dislocation neutral.

If we have a gas of dislocations in the positions fixed by
disorder and their interaction, a single charge can only spread to
the finite distance, $R$. It can be estimated in the way we did
for residual randomly distributed acceptors. If we assume that
dislocation are fixed in space by fluctuating donor potential the
extra electron charge spreads optimizing the sum of its Coulomb
energy and the energy of the shear deformation. Optimization leads
to $R\sim s$. Thus, the Coulomb potential of charges inside the
pinned Wigner crystal is valid at distances in the plane $r \gg
s$.

This system is clearly similar to the Coulomb glass and therefore
has the Coulomb gap. Indeed, as we mentioned above the derivation
of the Coulomb gap is based upon the observation that when a
localized electron is transferred to another localized state one
should take into account its 1/r Coulomb attraction with the hole
it has left (the excitonic effect). We claimed above that that the
Coulomb interaction is valid if $r \gg s$. Therefore, the linear
in energy Coulomb gap appears at the Fermi level. The width of
this gap is $e^{2}/\kappa R = e^{2}/\kappa s$. At $s \gg a$ the
Coulomb gap occupies only small fraction of the energy range
between interstitial and vacancy peaks. Away from the Coulomb gap
the density of states is almost constant. Thus, in the case of
intentional $\delta$-doping by remote donors we again arrive to
the Coulomb gap of density of states and correspondingly to ES law
at low temperatures.

Let us discuss the value of localization length $\xi$ in
Eq.~(\ref{TES}). In classical Wigner crystal $\xi \sim
a/r_s^{1/2}$. This is a quite small value which leads to too large
$T_0$ and very large resistances in the range of ES law. However,
close to the melting point of the Wigner crystal $\xi$ can be much
larger, making an observation of ES law in pinned Wigner crystal
more realistic. Magnetoresistance, observed in
Ref.~\onlinecite{Noh} may be related to ES variable range hopping
conductivity.

In conclusion, I emphasize again that I am dealing with the Wigner
crystal which in the absence of impurities slides on the positive
background. Impurities pin this Wigner crystal and lead to ES VRH
conductivity due to hops between its pieces. This situation is
similar to what happens in a system of many quantum dots situated
in random electrostatic potential of stray charges or in a system
of many electron puddles with random positive charge of their
background. Similar physics was recently theoretically studied in
quasi-one-dimensional systems~\cite{Fogler}.

I am grateful to M. M. Fogler, A. L. Efros and S. Teber for many
important discussions. This paper is supported by NSF DMR-9985785.


\end{document}